\documentclass[12pt]{article}
\usepackage{amssymb}
\usepackage{cite}
\usepackage{epsf}

\def\action{{\cal S}}
\def\courbure{{\cal R}}

\def\hmu{{\hat \mu}}
\def\hnu{{\hat \nu}}
\def\hsigma{{\hat \sigma}}

\def\sgn{\mbox{sgn}}
\newcommand{\Reel}{{\mathbb{R}}}

\newcommand{\nc}{\newcommand}
\nc{\beq}{\begin{equation}}
\nc{\eeq}{\end{equation}}
\nc{\beqa}{\begin{eqnarray}}
\nc{\eeqa}{\end{eqnarray}}
\nc{\lra}{\leftrightarrow}
\def\sfrac#1#2{{\textstyle\frac#1#2}}
\nc{\sss}{\scriptscriptstyle}
{\nc{\lsim}{\mbox{\raisebox{-.6ex}{~$\stackrel{<}{\sim}$~}}}
{\nc{\gsim}{\mbox{\raisebox{-.6ex}{~$\stackrel{>}{\sim}$~}}}

\newcommand{\pa}{a^{\prime}}
\newcommand{\pb}{b^{\prime}}

\newcommand{\ppa}{a^{\prime \prime}}

\newcommand{\fpa}{\frac{\pa}{a}}
\newcommand{\fpb}{\frac{\pb}{b}}

\newcommand{\fppa}{\frac{\ppa}{a}}

\def\diff#1#2{ \frac{\partial #1}{\partial #2 }}

\begin{document}

\begin{titlepage}
\begin{flushright}
LBNL 44356\\
UCB-PTH-99/49\\
McGill 99-32\\
Saclay t99/112 \\
hep-th/9910081 \\
\end{flushright}

\vskip.5cm
\begin{center}
{\huge{\bf Supergravity Inspired \\
Warped Compactifications \\
and \\ \vskip0.4cm
Effective Cosmological Constants}}
\end{center}
\vskip1.5cm

\centerline{ C. Grojean $^{a,b}$, J. Cline $^{c}$, and G. Servant $^{d}$  }
\vskip 15pt

\centerline{$^{a}$ Department of Physics, 
University of California, Berkeley, CA 94720}
\vskip 3pt
\centerline{$^{b}$ Theoretical Physics Group,
 Lawrence Berkeley National Laboratory, Berkeley, CA 94720}
\vskip 3pt
\centerline{$^{c}$ Physics Department, McGill University, 
Montr\'eal, Qu\'ebec, Canada H3A 2T8}
\vskip 3pt
\centerline{$^{d}$ CEA-SACLAY, Service de Physique Th\'eorique, 
F-91191 Gif-sur-Yvette, France}

\vglue .5truecm

\begin{abstract}
\vskip 3pt

\noindent
We propose a supergravity inspired derivation of a Randall--Sundrum's type action as an
effective description of the dynamics of a brane coupled to the bulk
through gravity only. The cosmological constants in the bulk and on the
brane appear at the classical level when solving the
equations of motion describing the bosonic sector of supergravities in ten and eleven
dimensions coupled to the brane.
They are related to physical quantities like the brane
electric charge and thus inherit some of their physical properties.
The most appealing property is their quantization:  in $d_\perp$ extra
dimensions, $\Lambda_{\mbox{\tiny brane}}$ goes like $N$ and
$\Lambda_{\mbox{\tiny bulk}}$ like $N^{2/(2-d_\perp)}$.  This dynamical
origin also explains the apparent fine-tuning required in the
Randall--Sundrum scenario.  In our approach, the cosmological constants
are derived parameters and cannot be chosen arbitrarily; instead they are
determined by the underlying Lagrangian.
Some of the branes we construct that support cosmological constant in the bulk
have supersymmetric properties: D3-branes of type
{\it IIB} superstring theory provide an explicit example.

\end{abstract}

\end{titlepage}

\section{Introduction}

The coexistence of two hierarchical scales in particle physics is
probably the most challenging puzzle to solve before hoping to
construct a quantum theory of gravity. When the Schwarzchild radius
($R_{\mbox{\tiny Sch}} = 2 {\cal G}_{N} m/c^2$) of a system of mass
$m$ becomes of the same order as its Compton length
($\lambda_C=\hbar/mc$), a quantum mechanical extension of general relativity
is surely needed.   Therefore the natural scale of quantum gravity
is the
Planck mass, $\sqrt{\hbar c^5/{\cal G}_N} \sim 10^{19}$
GeV. Understanding how, in such a theory, the tiny electroweak scale
observed in experimental particle physics can arise and be
stabilized against radiative corrections constitutes the so-called `gauge
hierarchy problem'.  In low energy supersymmetry \cite{LowSusy}, this
vast disparity in scales can be protected from quantum destabilization.
However a more fundamental explanation is certainly to be found in
string theory and its latest developments. String theory relates the
string scale to two other fundamental scales, namely the GUT scale
connected to gauge interactions, and the Planck scale connected to
the gravitational interaction. The link between these two is 
the geometry of extra dimensions, which can lower both scales
\cite{LowScales} down to the TeV range \cite{LargeDim} and thus
partially answer the gauge hierarchy problem, or at least translate
it into geometrical terms.

Subsequent to studies of thin shells in general relativity \cite{Shell}
and their revival in a $M$-theory context \cite{Ovrut,ChRe,BDL},
Randall and Sundrum (RS) have recently proposed \cite{RS} a new
phenomenological mechanism for solving the gauge hierarchy problem,
without requiring the extra dimension to be particularly large or
small--in fact it could be noncompact. An exponential hierarchy is
generated by the localization of gravity near a self-gravitating brane
with positive tension, obtained by solving Einstein equations. The
solution is a nonfactorizable metric, {\it i.e.,} a metric with an
exponentially decaying warp factor \cite{warp} along the
single extra dimension.  Restricting the Standard Model to a second
parallel brane with negative tension at some distance in this
transverse dimension, the electroweak scale in our world then follows
from a redshifting of the Planck scale on the second brane.  Since the
exponential suppression by the redshift factor does not require an
unnaturally large interbrane separation, the hierarchy problem can be
explained without fine tuning, and without requiring any special size
for the extra dimensions.

The cosmological implications of this scenario have been studied
\cite{cosmo,CGS}, with emphasis on the danger of placing the Standard Model
on a brane with negative tension since, for instance, the Friedmann
equation governing the expansion of the universe appears with a wrong
sign.  A similar difficulty is also faced \cite{RSunification} when
trying to reproduce the unification of gauge couplings.  The original
scenario can be modified \cite{RS,RSunification,RL} by maximizing the
warp factor on the Standard Model brane, which can be achieved if its
tension is taken positive. The two former problems are overcome but the
electroweak scale seems now difficult to accommodate.
More recently
it has been shown that the correct cosmological expansion can be
obtained if the second brane tension is negative, but not too much so
\cite{CGS2}.  
Thus the RS scenario remains attractive, especially with
regard to the possibility of an infinite extra dimension probed only by
gravity.  It is appealing that, despite a continuous Kaluza--Klein
spectrum without any mass gap, Newton's law of gravity is still
reproduced \cite{RS,RL,ADDK} within the current experimental
precision. Ref. \cite{NewtonLaw} also proposed explicit
models where a mass gap separates the `massless graviton' from its
KK excitations while the Yukawa type deviations from the 4D Newton law
remain compatible with experimental bounds.

Although the gravity localization mechanism seems to be specific to
codimension one branes, several works \cite{ADDK,Junctions} have managed
to extend it by considering many intersecting codimension one
branes.\footnote{See also ref. \cite{ChPz} for a recent construction of
warped compactification in two transverse dimensions.} Oda and Hatanaka
{\it et al.} \cite{manybranes} also obtain solutions with a more involved
content of branes with a single one extra-dimension.  In this context also,
ref. \cite{CGS2} finds a cosmological solution
for the bulk and the branes inflating at the same rate with a time
dependant Planck mass. In that case, the hierarchy between the weak and
the Planck scales is fixed at the end of inflation.
.

Undoubtedly, the localization of gravity by the RS mechanism has rich
phenomenological and cosmological consequences 
\cite{cosmo,CGS,RL,RSunification,Goldberger,RScosmo,RSpheno,CGS2,manybranes,ADDK};
but at the present stage it seems
lacking in generality, and it suffers from apparently {\it ad hoc}
fine-tunings required between the cosmological constants in the bulk and
on the branes, in order to obtain a solution to Einstein equations. 
Verlinde \cite{Verlinde} has reexamined the RS scenario in superstring
language and shown that the warp factor can be interpreted as a
renormalization group scaling.  In the context of the
 AdS/CFT correspondence,  the extra dimension plays the role of the
energy scale.

In this paper, we offer a derivation of the effective action used by
RS, starting from a more fundamental, string-inspired origin.  Recent
works \cite{Ovrut,BDL,sugrawalls,Cvetic} have studied the dynamics of a
brane-universe; here we propose an explicit embedding of
the RS model in supergravity theories and examine its physical
implications, following refs.\ \cite{NewtonLaw,Kehagias}, which
have previously addressed this question at a more formal level.
Our starting point will be the bosonic action of
supergravity theories in ten or eleven dimensions. We emphasize that,
instead of neglecting  various fields specific to these actions like
the dilaton and some $n$-differential forms, taking them into account
can lead to an effective description in terms of cosmological
constants. Using $p$-brane solutions\footnote{
The branes we construct are solution to the {\it bosonic} equations of motion
and thus, as we will see later, they are not necessarily supersymmetric even if
they are embedded in a supergravity theory.}, we construct such a description
for codimension one branes, which allows us to identify the effective
cosmological constants with physical quantities like the electric
charge carried by the brane and its mass density on the worldvolume.
Since the electric charge of a $p$-brane obeys a generalized Dirac
quantization rule, we are led to the interesting conclusion that the
cosmological constants are also quantized. 

The advantage of this approach is that we derive the stress-energy
tensor $T_{\hmu\hnu}$, which is needed to solve the Einstein equations,
starting from an action for fundamental fields, rather than putting it
in by hand.  Thus our $T_{\hmu\hnu}$ is on the same footing as the
Einstein tensor itself, from the point of view of fundamentality.
Moreover we are able to generalize the
procedure to higher codimension brane-universes ({\it e.g.,} 3-branes
embedded in more than one extra dimension), providing some of the first
such solutions.  In this case the bulk energy is no longer a
cosmological ``constant,'' but depends on the distance from the brane.

\section{Brane cosmological constant as a warp in an anti-de Sitter bulk}

We begin with a review  of the model studied by Randall and Sundrum
\cite{RS}.
This model is a particular case of the ones proposed by Chamblin
and Reall \cite{ChRe}, in which a scalar field was coupling a
dynamical brane to an embedding bulk.
Here we consider the restricted scenario of a static brane embedded in a
spacetime curved by a bulk cosmological constant
$\Lambda_{bk}$. The physics of this model is governed by the following action:
\beq
	\label{RSaction}
\action_{RS} = \int d^{p+1} x\, d^{d_\perp}y \, \sqrt{|g|}\,
\left( \frac{\courbure}{2\kappa^2} - \Lambda_{bk} -
\Lambda_{br} \delta^{d_\perp} (\sqrt{|g_\perp|}\, y)
\right) \ ,
\eeq
where $y^I=0$ is the location of the brane in the transverse (extra
dimensional) subspace
and $g_\perp$ is the determinant of the metric, assumed to be
factorizable, in this subspace.
The Einstein equations derived from (\ref{RSaction})
 when the transverse space is flat
are
(Greek indices denote longitudinal coordinates,
$\mu=0\ldots p-1$ and Latin indices are  coordinates transverse 
to the brane,
$I=1\ldots d_\perp$):
\begin{eqnarray}
G_{\mu\nu} & = &  -\kappa^2 \left(
\Lambda_{bk} + \Lambda_{br} \delta^{d_\perp} (\sqrt{|g_\perp|}\, y) \right)\,
g_{\mu\nu} \ ;\\
G_{IJ} & = & -\kappa^2 \Lambda_{bk} \, g_{IJ} \ .
\end{eqnarray}
Randall and Sundrum solved these equations in the case of a codimension one brane.
With the ansatz
\begin{equation}
ds^2 = a^2 (y)\, dx^\mu \otimes dx^\nu \eta_{\mu\nu} + b^2 (y)\, dy \otimes dy  \ ,
\end{equation}
the Einstein equations reduce to 
\begin{eqnarray}
\label{appeq}
& \displaystyle
p \fppa + \frac{p(p-1)}{2} \left( \fpa \right)^2
- p \fpa \fpb =
-\kappa^2 \left( \Lambda_{bk} + \Lambda_{br} \, \delta (|b|y) \right) b^2
\ ;\\
& \displaystyle
\frac{p(p+1)}{2} \left( \fpa \right)^2
=
- \kappa^2 \Lambda_{bk} \, b^2
\ ,
\end{eqnarray}
where  primes denote derivatives with respect to
the transverse coordinate $y$.
For this system of equations to admit a solution that matches the singular terms,
a fine-tuning between $\Lambda_{bk}$ and $\Lambda_{br}$ is necessary:
\beq
	\label{eq:finetuning}
\Lambda_{bk} = - \frac{p+1}{8p} \, \kappa^2 \Lambda_{br}^2
\ .
\eeq
A general solution then takes the form:
\beq
a(y) = f(|y|) \ \ \
\mbox{and } \ \ \
b(y) = {\cal N} \frac{f'(|y|)}{f(|y|)} \ ,
\eeq
where $f$ is a regular function and the constant ${\cal N}$
is related to the brane cosmological constant by:
$|{\cal N}|=-2p \epsilon/(\kappa^2 \Lambda_{br})$, $\epsilon$ being the sign of $f'(0)/f(0)$.
A particular class of solutions  that will play an important role in our analysis
corresponds to:
\beq
	\label{eq:solus}
a(y) = \left( l + |y|/R \right)^{n_a} \ \ \
\mbox{and } \ \ \
b(y) = \frac{n_a {\cal N} R^{-1}}{l+ |y|/R} \ ,
\eeq
where $R$ and $l$ are two positive constants.
An appropriate change of coordinates brings this solution to the form proposed by
Randall and Sundrum \cite{RS}: defining $X^\mu = l^{n_a} x^\mu $ and
$Y= \sgn(y) n_a {\cal N} \ln (1+|y|/(Rl))$, the metric reads:
\beq
ds^2 = e^{2 \,{\rm sgn}(n_a) |Y/{\cal N}|} \, dX^2 + dY^2
\ .
\eeq

If the brane located at the origin is identified as the ``Planck brane''
of Lykken--Randall \cite{RL}, an electroweak scale will be generated on
the ``TeV brane'' if and only if the power $n_a$ is negative, which
corresponds to a positive cosmological constant on the Planck
brane.\footnote{This connection between the signs of $n_a$ and $\Lambda_{br}$
is specific to one transverse dimension. In section 4, we will see that we
can have $n_a>0$ whereas $\Lambda_{br} >0$. In any case, the discussion about
the hierarchy problem deals with the sign of $n_a$ only.}  Another
motivation for requiring $n_a<0$ comes from computing the four-dimensional
effective Planck mass, $M^2_{Pl}=M^3 \int dy\, a^2 |b|$, which is finite
for $n_a<0$ but diverges for $n_a>0$.

\vskip 0.5cm
\centerline{\epsfxsize=3.4in\epsfbox{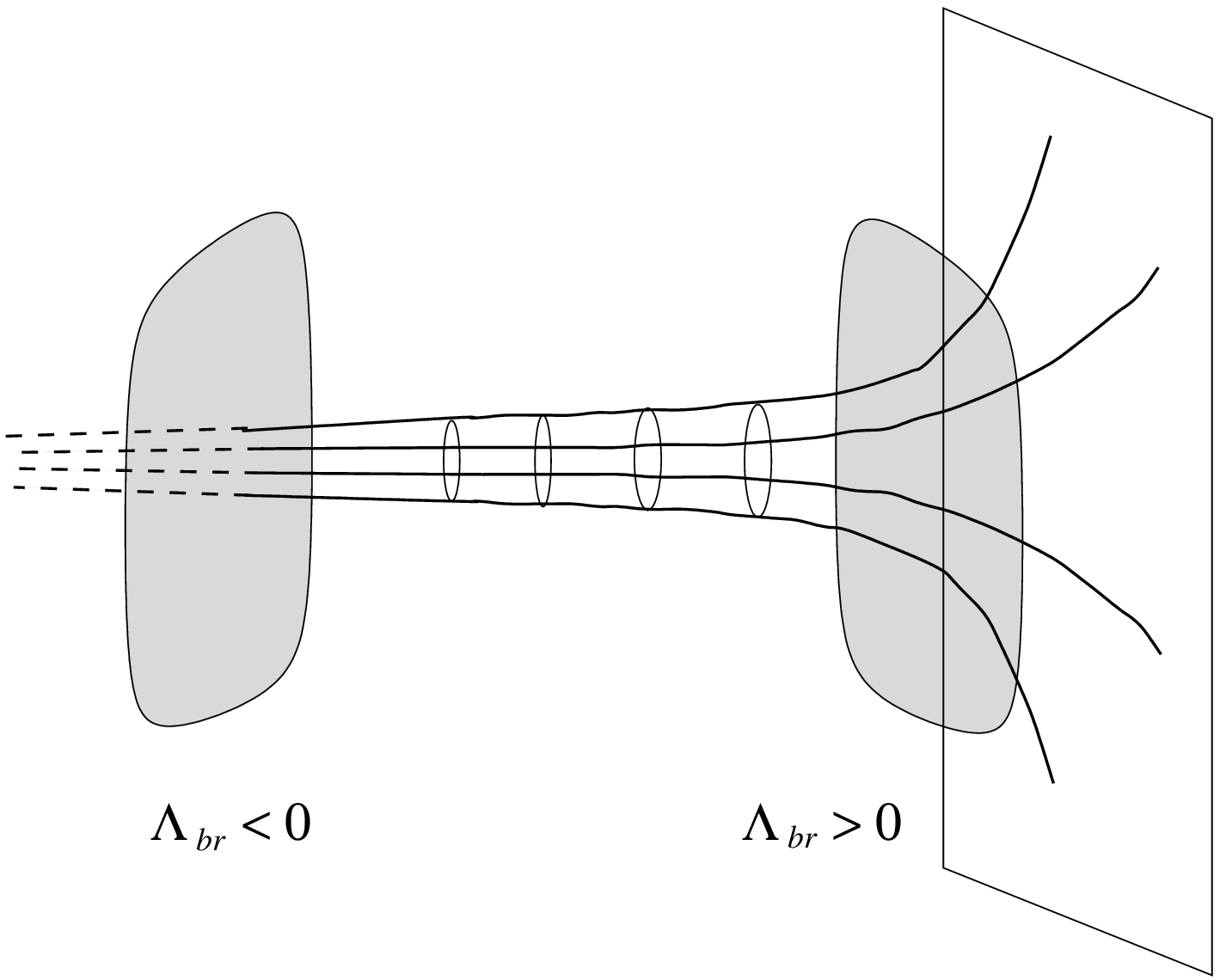}}
\noindent {\small Figure 1. The boundary of an anti-de Sitter of dimension $p+2$
space is topologically
$S^1\times S^{p}$. In the system of coordinates $x^\mu$ and $r$, this boundary is
located at $r=0$ and $r=\infty$:
the piece at infinity is a $p+1$-dimensional Minkowskian space,
while the horizon at $r=0$ corresponds
to the union of a point and $\Reel \times S^{p-1}$.
A codimension one brane embedded in this $AdS$ space acts as a warp
in the sense that it cuts a part of the bulk:
a brane with a positive cosmological constant cuts the vicinity of
the boundary located at the infinity, while a brane with a negative cosmological constant
removes the horizon at the origin.
}
\vskip 0.5cm

We can make another diffeomorphism that clarifies the geometry of the solution.
Defining $r=R_0 (l+|y|/R)^{n_a}$, with $R_0=|{\cal N}|$, we now obtain:
\beq
ds^2 = \left( \frac{r}{R_0} \right)^2 dx^2 + \left( \frac{R_0}{r} \right)^2 dr^2
\ ,
\eeq
where we see that the geometry of the bulk corresponds to an anti-de Sitter space
of radius $R_0$, or at least a slice of an anti-de Sitter space, since the variable
$r$ ranges only over a part of $\Reel$.
Indeed, for $n_a>0$, the range of variation of $r$ is restricted to
$[l^{n_a}, +\infty )$, while for $n_a<0$ this range becomes
$[0,l^{n_a}]$. Although in both cases the whole $AdS$ space is covered
in the limit $l \to 0$, it is interesting to note which part
is cut when $l \not =0$. As we will argue in the appendix, the boundary
of an anti-de Sitter space of dimension $p+2$
space is topologically
$S^1\times S^{p}$, and in the system of coordinates $x^\mu$ and $r$, this boundary is
located at $r=0$ and $r=\infty$:
the piece at infinity is a $p+1$-dimensional Minkowskian space,
while the horizon at $r=0$ corresponds
to the union of a point and $\Reel \times S^{p}$. So
the $n_a<0$ case, which corresponds to a positive cosmological constant $\Lambda_{br}$
on the brane, removes the part at infinity, while
the $n_a>0$ case, {\it i.e.} $\Lambda_{br}<0$, cuts the horizon at the origin.
Note that in the AdS/CFT correspondence \cite{Mal}, a superconformal theory
describes the dynamics of a brane near the horizon of an $AdS$ space
while this dynamics should become free near infinity \cite{GM}.

As presented, the model studied by Randall and Sundrum leaves one
wondering whether it can be derived from some more fundamental starting
point.  In particular, the {\it ad hoc} fine-tuning between the
cosmological constants is rather mysterious and begs for a better
understanding.  One suggestion is that this relation might arise from
the requirement that tadpole amplitudes are zero in the underlying
string theory \cite{CGS}.   (See also ref.\ \cite{Ellwanger} for recent
progress about this question).  Here we will see the cosmological
constants as effective parameters which cannot be chosen arbitrarily,
so the fine-tuning problem is ameliorated.  The aim of this
work is to motivate the RS model from a supersymmetry/superstring
framework.

\section{Effective cosmological constants from dynamics of codimension one branes}

In this section, we would like to show that the theory derived from
the action (\ref{RSaction}) can be seen as an effective description of a
brane of codimension one, {\it i.e.,} of an extended object with
$p$ spatial dimensions embedded in a ($p+2$) dimensional spacetime.

The dynamics of an object extended in $p$ spatial directions is 
governed by the generalization of the Nambu--Goto action\footnote{Concerning the
indices, our conventions will be the following:
hatted Greek indices are spacetime indices ($\hmu=0\ldots D-1$)
while Latin indices are worldvolume indices($a=0\ldots p$).}  \cite{NambuGotto}:
\begin{equation}
\action_{NG} = - M^{p+1}_b \int d^{p+1} \xi
\sqrt{\left| 
\det \left( \diff{X^\hmu}{\xi^a} \diff{X^\hnu}{\xi^b} g_{\hmu\hnu}\right)
\right|}
\ ,
\end{equation}
where $X^{\hmu} (\xi^a)$ are the coordinates in the embedding spacetime 
of a point on the brane characterized by its worldvolume coordinates
$\xi^a$; $M_b$ is the scale mass in so-called ``$p$-brane units'' which is
simply related to the Planck scale, $M$, in the embedding spacetime; see below
eq.\ (\ref{betapeq}).
This action is known \cite{Polyakov} to be equivalent to:
\begin{equation}
	\label{eq:PolyakovPbrane}
\action_{P} = M^{p+1}_b \int d^{p+1} \xi\,
\left(
-\frac{1}{2}
\sqrt{|\gamma|}
\gamma^{ab}
\partial_a X^\hmu \partial_b X_\hmu
+\frac{p-1}{2} \, \sqrt{|\gamma|}
\right)
\ ,
\end{equation}
where  $\gamma_{ab}$ is an auxiliary field that gives the metric on 
the worldvolume.

Superbranes have been constructed \cite{Branes} as classical solutions of
supergravity theories in ten or eleven dimensions: they are BPS
objects, since they preserve half of the supersymmetries; they have a
Poincar\'e invariance on their worldvolume universe and also a rotational
invariance in the transverse space.
A $p$-brane is therefore coupled to the low-energy effective theory of superstrings.
Below the fundamental energy scale, identified as the energy of the first massive
excitations of the string, the theory can be described by supergravity theories
whose bosonic spectrum contains the metric, a scalar field (the dilaton)
and numerous differential forms.
The bosonic effective action, in supergravity units, takes the general form
($\kappa^2=M^{2-D}$):
\begin{equation}
	\label{eq:braneLagrangienEffectif}
\action_{eff} = \int d^Dx \, \sqrt{|g|} 
\left(
\frac{1}{2\kappa^2}\, \courbure 
-\frac{1}{2}\, \partial_\hmu \Phi \partial^\hmu \Phi
-\frac{1}{(p+2)!}\, e^{\alpha_{p} \Phi}
F_{\hsigma_1\ldots \hsigma_{p+2}}  F^{\hsigma_1\ldots \hsigma_{p+2}}
\right)
\ ,
\end{equation}
where $ F_{\hmu_1\ldots\hmu_{p+2}} = (p+2) \,
	\partial_{[\hmu_1}\, A_{\hmu_2\ldots \hmu_{p+2}]} $ is the
field strength of the ($p+1$)-differential form $A$, whose coupling
to the dilaton is measured by the coefficient $\alpha_p$.
The coefficient $\alpha_p$ is explicitly
determined by a string computation:
the coupling of the dilaton to differential forms
from the Ramond-Ramond sector
appears at one loop and thus
$\alpha_p^{RR}/\sqrt{2\kappa^2}=(3-p)/2$ in supergravity units,
while the Neveu-Schwarz--Neveu-Schwarz two-form couples at tree level, so
$\alpha_{1}^{NS}/\sqrt{2\kappa^2}=-1$.
In some cases, we can also add a Chern--Simons term
($A\wedge F \wedge F$) to the action, but
it does not have any effect on the classical solutions.

The $p$-brane couples to a ($p+1$)-differential form, which
results in the addition of a Wess--Zumino term to
the free action (\ref{eq:PolyakovPbrane}):
\begin{eqnarray}
	\nonumber
\action_{P} = M^{p+1} \int d^{p+1} \xi\,
\left(
-\frac{1}{2}
\sqrt{|\gamma|}
\gamma^{ab}
\partial_a X^\hmu \partial_b X^\hnu
g_{\hmu\hnu} (X) \, e^{\beta_p \Phi}
+\frac{p-1}{2} \, \sqrt{|\gamma|}
\right. \\
	\label{eq:actionbraneunivers}
\left.
+\frac{{\cal A}_{WZ}}{(p+1)!}\, \epsilon^{a_1\ldots a_{p+1}}
\, \partial_{a_1} X^{\hmu_1} \ldots \partial_{a_{p+1}} X^{\hmu_{p+1}}
A_{\hmu_1\ldots \hmu_{p+1}}
\right)
\ .
\end{eqnarray}
The functions  $g_{\hmu\hnu}$ and $e^{\beta_p \Phi}$ implicitly depend 
on worldvolume coordinates $\xi$ through
their dependence in the embedding coordinates $X$.
The coefficient $\beta_p$ defines the ``$p$-brane units;'' it is fixed
\cite{ReviewBranes} by requiring the same scaling behavior for $\action_{eff}$
and $\action_{P}$, which leads to
\begin{equation}
\label{betapeq}
\beta_p = -\frac{\alpha_p}{p+1}
\ .
\end{equation}
The relation between $M_b$ and $M$ then follows from the value of this coupling to the dilaton:
$M_b=e^{\beta \phi_\infty/(p+1)} M$, $\phi_\infty$ being the vacuum expectation value
of the dilaton.

To proceed, we must now relax some of the constraints imposed by
supersymmetry, while still maintaining the form of the action.  For
example in string theories, the values of $p$ and $D$ are related to
one another in order to have supersymmetry on the worldvolume universe
\cite{SuperBranes}.  Also, as just mentioned, the coupling to the
dilaton is fixed.  By relaxing these constraints, we give up any claim
that the following construction is a direct consequence of string
theory.  On the other hand it might be hoped that our results will
persist in a realistic low energy limit of string theory, which
includes the effects of supersymmetry breaking.  In what follows, we
will elucidate how the various fields, which play a crucial role for
the existence of branes in supergravity, can give rise to an effective
stress-energy tensor which resembles the cosmological constant terms
needed for the Randall--Sundrum scenario.

The equations of motion derived from $\action_{eff}+\action_P$ are
\begin{eqnarray}
\nonumber 
&&   
G_{\hmu\hnu} = 
\kappa^2\, \partial_\hmu \Phi \partial_\hnu \Phi
+\frac{2\kappa^2}{(p+1)!} \, e^{\alpha_p \Phi}\, 
	F_{\hmu\hsigma_1\ldots\hsigma_{p+1}} 
	F_{\hnu}{}^{\hsigma_1\ldots\hsigma_{p+1}} 
\\
&&	\label{eq:braneRicci}  \rule{1.5cm}{0pt} 
+\frac{1}{2} \left( 
-\kappa^2 \, \partial_\hsigma \Phi \partial^\hsigma \Phi
-\frac{2 \kappa^2 }{(p+2)!} \, e^{\alpha_p \Phi}\,
     F_{\hsigma_1\ldots\hsigma_{p+2}}  F^{\hsigma_1\ldots\hsigma_{p+2}}
\right)\, g_{\hmu\hnu}\,
+ T_{\hmu\hnu} 
\ ;
\\
	 \label{eq:branedilaton} 
& &
D_\hmu D^\hmu \Phi = \frac{\alpha_p }{(p+2)!}\, e^{\alpha_p \Phi}\, 
F_{\hsigma_1\ldots\hsigma_{p+2}} F^{\hsigma_1\ldots\hsigma_{p+2}}
+ T_\Phi
\ ;
\\
	\label{eq:braneTenseur}
& &
\partial_{\hmu_0} \left( \sqrt{|g|} \, e^{\alpha_p \Phi}\, 
F^{\hmu_0\ldots\hmu_{p+1}}
\right) = J^{\hmu_1\ldots\hmu_{p+1}}
\ ;
\\
	\label{eq:braneMetriqueInduite}
& &
\gamma_{ab} = \partial_a X^\hmu \partial_b X^\hnu g_{\hmu\hnu} \, 
e^{\beta_p \Phi}
\ ;
\\
	\nonumber
& &
\partial_a \left( \sqrt{| \gamma |} \gamma^{ab} \partial_b X^\hnu
g_{\hmu\hnu} e^{\beta_p \Phi} \right) = 
\frac{1}{2}\, \sqrt{| \gamma |} \gamma^{ab} \partial_a X^{\hsigma_1}
\partial_b X^{\hsigma_2} 
\partial_\hmu \left( g_{\hsigma_1\hsigma_2} e^{\beta_p \Phi} \right)
\\
&&	\label{eq:braneCoordonnees}  \rule{3.5cm}{0pt}
-\frac{{\cal A}_{WZ}}{(p+1)!}\, \epsilon^{a_1\ldots a_{p+1}} \
\partial_{a_1} X^{\hsigma_1} \ldots \partial_{a_{p+1}} X^{\hsigma_{p+1}} \,
F_{\hmu \hsigma_1 \ldots \hsigma_{p+1}}
\ .
\end{eqnarray}
The stress-energy tensor $T_{\hmu\hnu}$ of the brane is given by
\begin{equation}
	\label{eq:stressenergy}
T_{\hmu\hnu} = - \kappa^2 M^{p+1} \int d^{p+1} \xi
\, \sqrt{|\gamma|}\, \gamma^{ab} 
\partial_a X^{\hmu^\prime}  \partial_b X^{\hnu^\prime} \,
g_{\hmu^\prime\hmu} g_{\hnu^\prime\hnu} \,
e^{\beta_p \Phi}\, \frac{\delta^D (x-X(\xi))}{\sqrt{|g|}}
\ .
\end{equation}
The electric current created by the brane is
\begin{eqnarray}
	\nonumber
J^{\hmu_1\ldots\hmu_{p+1}} = 
-\frac{{\cal A}_{WZ}}{2} \, M^{p+1} \int d^{p+1} \xi
\, \epsilon^{a_1\ldots a_{p+1}} \ \ \ \ \\
	\label{eq:electriccurrent}
\partial_{a_1} X^{\hmu_1} \ldots \partial_{a_{p+1}} X^{\hmu_{p+1}}
\, \delta^D (x-X(\xi))
\ .
\end{eqnarray}
And the source current for the dilaton equation is
\begin{equation}
T_\Phi = \frac{\beta_p\, M^{p+1}}{2} \int d^{p+1} \xi
\, \sqrt{|\gamma|}\, \gamma^{ab} \partial_a X^{\hmu}  \partial_b X^{\hnu} \,
g_{\hmu\hnu}\, 
e^{\beta_p \Phi}\, \frac{\delta^D (x-X(\xi))}{\sqrt{|g|}}
\ .
\end{equation}

We will solve these equations in the case of a codimension one brane
and we will see in the next section how the analysis can be extended to higher codimension.
First we choose a system of spacetime coordinates related to the brane:
\begin{eqnarray*}
&& \mbox{worldvolume coordinates: } \ \  x^\mu \ \ \ \ \mu=0\ldots p \ ;\\
&& \mbox{transverse coordinate: }  \ \ y  \ ,
\end{eqnarray*}
in the physical gauge where $X^\mu (\xi) = \xi^\mu$.

We are looking for a solution with a Poincar\'e invariance in 
($p+1$) dimensions, so that we can make the following ansatz for the metric:
\begin{equation}
	\label{eq:ansatzMetric}
ds^2 = 
e^{2 A(y)} dx^\mu \otimes dx^\nu \eta_{\mu\nu}
+e^{2 B(y)} dy \otimes dy   \ .
\end{equation}
The nonvanishing components of the ($p+1$)-differential form that couples
to the $p$-brane are
\begin{equation}
	\label{eq:ansatzA}
A_{\mu_1\ldots \mu_{p+1}} =
- \epsilon_{\mu_1 \ldots \mu_{p+1}}
\frac{1}{{\cal A}_{WZ}}
\ e^{C(y)}
\ ,
\end{equation}
where $\epsilon_{\mu_1 \ldots \mu_{p+1}}$ is the antisymmetric
tensor normalized to $\pm 1$.

It is well known that (see for instance \cite{ReviewBranes} for a review),
corresponding to the ansatz (\ref{eq:ansatzMetric}--\ref{eq:ansatzA}),
the solutions of eqs (\ref{eq:braneRicci}--\ref{eq:braneCoordonnees})
can be expressed in terms of
a harmonic function $H(y)$:
\begin{eqnarray}
	\label{eq:metriquebrane}
&&
ds^2  =
H^{2 n_x} \, dx^\mu \otimes dx^\nu \, \eta_{\mu\nu}
+ H^{2 n_y} \, dy \otimes dy  \ ;
\\
&&
e^{\Phi}  =
H^{n_\Phi} \ e^{\phi_\infty} \ \ \ 
(\phi_\infty \ \mbox{ is the value of $\Phi$ at infinity}) \ ;
\\
	\label{eq:tenseurelectrique}
&&
F_{y\mu_1\ldots\mu_{p+1}} =  
\epsilon_{\mu_1 \ldots \mu_{p+1}}
\frac{1}{{\cal A}_{WZ}}\,
e^{-\alpha_p \phi_\infty /2}\, \frac{d H^{-1}}{dy} \ ;
\end{eqnarray}
where the powers are given by
\begin{equation}
	\label{eq:powers}
n_x = \frac{2 \kappa^2}{p\, {\cal A}_{WZ}^2}
\ \ \ \
n_y =   \frac{2 (p+1) \kappa^2}{p\, {\cal A}_{WZ}^2}
\ \ \ \
n_\Phi = \frac{\alpha_p}{{\cal A}_{WZ}^2} \ .
\end{equation}
The consistency of the whole set of equations of motion with our $p$-brane ansatz
requires to adjust the coefficient of the Wess--Zumino term to
to the coupling to the dilaton by
\begin{equation}
	\label{eq:WZ}
{\cal A}_{WZ}^2 = -2 \kappa^2 \frac{p+1}{p} + \frac{\alpha_p^2}{2} \ ,
\end{equation}
the whole set of equations of motion is now equivalent to Poisson's equation,
\begin{equation}
\frac{d^2 H}{dy^2} = -\sfrac12 {\cal A}_{WZ}^2 M^{p+1}
e^{-\alpha_p \phi_\infty /2}\ \delta (y) \ ,
\end{equation}
the solution of which reads
\begin{equation}
	\label{eq:solH}
H(y) = l -\sfrac14 {\cal A}_{WZ}^2 M^{p+1} e^{-\alpha_p \phi_\infty /2}\ |y|
\ ,
\end{equation}
where $l$ is an arbitrary positive constant that can be normalized to one if
a flat Minkowski space in the vicinity of the brane is wanted.
At this stage, it is worth noticing that the derivation follows directly from
the bosonic equations (\ref{eq:braneRicci})-(\ref{eq:braneCoordonnees})
and no supersymmetric argument has been used.
The full supergravity
equations also include a Killing spinor equation that can be consistently
solved, provided that the coupling of the differential form to the dilaton
takes its stringy value.  This promotes the bosonic solution to a BPS one.

It is interesting to substitute this solution back into the Einstein equations
(\ref{eq:braneRicci}) to obtain:
\begin{eqnarray}
	\nonumber
G^{\mu\nu} & = &
-\frac{\kappa^2}{{\cal A}_{WZ}^2}\,
\left( 1 - \frac{{\alpha_p^2}}{2\,{\cal A}_{WZ}^2 } \right)
H^{-2(n_y+1)} \,  (H^\prime)^2 g^{\mu\nu}
\\
&&  \rule{1cm}{0pt}
-\kappa^2 M^{p+1} \, H^{-(1+n_x(p+1))} e^{-\alpha_p \phi_\infty/2} \,
\frac{\delta (y)}{\sqrt{g_{yy}}} \, g^{\mu\nu}
\ ;
\label{eq:Einstein1}
\\
G^{yy} & = &
-\frac{\kappa^2}{{\cal A}_{WZ}^2}\,
\left( 1 + \frac{{\alpha_p^2}}{2\,{\cal A}_{WZ}^2 } \right)
H^{-2(n_y+1)} \,  (H^\prime)^2 \, g^{yy}
\ .
\label{eq:Einstein2}
\end{eqnarray}
In the limit of decoupling between the brane and the dilaton,
{\it i.e.,} $\alpha_p=0$, which also corresponds
to $n_y=(p+1)n_{x}=-1$ using the constraint (\ref{eq:WZ}),
the Einstein tensor involves two constants
$\Lambda_{bk}^\circ$ and $\Lambda_{br}^\circ$:
\begin{eqnarray}
G^{\mu\nu} & = &
-\kappa^2 \left(
\Lambda_{bk}^\circ + \Lambda_{br}^\circ \frac{\delta (y)}{\sqrt{g_{yy}}} \right)
 g^{\mu\nu}  \ ;
\\
G^{yy} & = &
-\kappa^2 \Lambda_{bk}^\circ \,  g^{yy}
\ .
\end{eqnarray}
If we keep the factors $\alpha_p\phi_\infty$ fixed (since $\phi_\infty$ could
go to infinity as $\alpha_p\to 0$), these constants are given by
\begin{equation}
\Lambda_{bk}^\circ = -\frac{p+1}{8p} M^{p+2}\, e^{-\alpha_p \phi_\infty}
\ \ \ \mbox{and } \ \ \
\Lambda_{br}^\circ = M^{p+1} \,e^{-\alpha_p \phi_\infty/2}
\ .
\end{equation}
They can be interpreted as effective cosmological constants since the metric
(\ref{eq:metriquebrane}) is a solution to the Einstein equations derived
from the RS action (\ref{RSaction}).

The expression of the cosmological constants in terms of fundamental Planck mass in
$D$ dimensions
may give some insight into the origin of the apparently {\it
ad hoc} fine-tuning (\ref{eq:finetuning}) of the RS mechanism:  here
the cosmological constants are no longer fundamental parameters and the
fine-tuning problem appears in a different way; in the present language
it is a consequence of taking the limit where the dilaton decouples
from the brane.  Of course this represents just one point in the full
parameter space.  The more general solution, when the dilaton does not
decouple, is a bulk energy density which depends on $y$, rather than a
cosmological constant term.
Regardless of this difference, one can
still obtain a rapidly  decaying warp factor, as long as $\alpha_p$
remains small enough -- the warp factor follows a power law whose
exponent is inversely proportional to $\alpha_p$.
The new insight, then, is that the original RS
solution is only the simplest possibility within a whole class of
solutions which can solve the hierarchy problem.

Furthermore, our approach links the energy densities of the brane and
bulk to physical quantities like the charge, $Q_e$, associated to the electric
current (\ref{eq:electriccurrent}):
\begin{equation}
Q_e = \int_{S^{D-p-2}} e^{\alpha_p \Phi}\, \star F
= \int_{S^{D-p-2}} \frac{e^{\alpha_p \phi_\infty}}{{\cal A}_{WZ}}
\diff{H}{y}
= \sfrac{1}{2} {\cal A}_{WZ}^2 M^{p+1}
\ .
\end{equation}
Not only is such a charge
conserved, but it also obeys Dirac's quantization rule
\cite{quantization}: solutions exist where the fiducial value of the
electric charge is multiplied by an integer and these can be
interpreted as a superposition of $N$ parallel branes.
From the multiplication of the source (\ref{eq:actionbraneunivers}) by a factor $N$,
we easily deduce the following scaling when disentangling the contribution
from the source and from the bulk
in (\ref{eq:Einstein1})--(\ref{eq:Einstein2})\footnote{It is important
to notice that the singular part of the Einstein tensor depends in the source
not only through the powers of $H$  but also intrinsically.}:
\begin{equation}
\Lambda_{bk} \sim Q_e^2 \ \ \ \
\mbox{and} \ \ \ \
\Lambda_{br} \sim Q_e
\ ,
\end{equation}
which assures that $\Lambda_{bk}$ and $\Lambda_{br}$ are quantized like
$N^2 \Lambda_{bk}^\circ$ and $N \Lambda_{br}^\circ$ respectively.

A serious shortcoming with the above solution is that the dilaton
decoupling regime requires a purely imaginary Wess--Zumino term (see
eq.\ (\ref{eq:WZ})), which implies an imaginary hence unphysical value
for the electric charge.  Let us consider what happens if we insist
that ${\cal A}_{WZ}$ be real-valued, which would happen if the dilaton
coupling, $\alpha_p$, was sufficiently large, instead of zero as we
previously assumed.  In this case eq.\ (\ref{eq:solH}) implies that the
harmonic form $H(y)$ vanishes at some value of $y$, $y_{\sss H} = l/c$,
where $c = M^{p+1}e^{-\alpha_p\phi_\infty/2} {\cal A}_{WZ}^2/4$.  The
exponents $n_x$ and $n_y$ are positive, so the metric coefficients
vanish at $y=y_{\sss H}$; this indicates the presence of a horizon
surrounding the brane.  The nontrivial dependence on $y$ of the brane
metric coefficient, $e^{2 A(y)}$, means that the compactification is
still warped; however it is not an exponential warp factor as in the
solution of Randall and Sundrum.  This can be seen by transforming to
the coordinate $d\bar y = e^{B(y)}dy$, which measures physical distance
in the bulk.  In this coordinate,
$e^{2 A(y)} = (l^{1+n_y} - c(n_y+1)\bar y)^{2n_x/(1+n_y)}$.

If we now imagine placing a visible sector brane (with such small
tension that it has negligible effect on the background geometry
\cite{RL}) very close to the horizon, physical masses on that brane
will be suppressed relative to the string scale by the small factor
$e^{2 A(y)}$.  However one must fine-tune the closeness of the brane to
the horizon to get particles of weak-scale masses, so this does not
provide a natural solution to the weak-scale hierarchy problem.

Because of the imaginary value of ${\cal A}_{WZ}$ required for the
solution which corresponds to that of Randall and Sundrum, our
construction is still just a tantalizing hint at a stringy origin for
their proposal.
To be more convincing, it is essential to overcome this problem.
In the next section, we will adress this issue by going to a higher
number of (still non copact) extra dimensions, in the space transverse
to the brane. However, it may happen that the problem of the imaginary
Wess--Zumino coupling also disappears when considering the compactification
of some of these extra dimensions,
requiring a more complete analysis involving some interacting moduli
fields in gauged supergravity theories\footnote{This question has been
recently addressed by Behrndt and Cveti{\v c} \cite{Cvetic}. See also
ref. \cite{Ovrut} for an earlier discussion.}.  The problem should also
be reconsidered in a more complicated version \cite{Romans} of ten
dimensional {\it IIA} supergravity including mass terms since a
codimension one supersymmetric object, the D-8 brane, has been
constructed by Bergshoeff {\it et al.} \cite{D8}.  This subject was
partially addressed in the recent references \cite{ChPeRe}.



In summary, our study of codimension one branes suggests that the
cosmological constants introduced by Randall and Sundrum are an
effective description of the dynamics of a more complicated set of
fields governing the physics of a brane that couples to the bulk
through gravitational interactions only.  Thus those effective
cosmological constants inherit some physical properties of the brane,
an intriguing one being their quantization.  We point out that, for
codimension one branes with no dilaton coupling, the solution
(\ref{eq:metriquebrane}) belongs to the general class of solutions
(\ref{eq:solus}).  Since the exponent $n_a=n_x=-1/(p+1)$ is negative,
it follows from the general discussion of section 2 that this field
configuration has an exponential decaying warp factor and thus can
solve the gauge hierarchy problem in the manner proposed by Lykken and
Randall \cite{RL}.  Namely, physical particle masses will be
exponentially suppressed on any test-brane (``TeV brane'') placed
sufficiently far from the ``Planck brane'' featured in our solution.

\section{Generalization to higher codimension brane-universe}

We would now like to generalize the previous results
to the case of a brane-universe of codimension greater
than one. Requiring  rotational invariance in the transverse space,
the ansatz for the metric and for the ($p+1$)-differential form will be
a function only 
of the distance $r$ in the transverse space:
\begin{equation}
r = \sqrt{y^I y^J \delta_{IJ}} \ .
\end{equation}
The solutions (\ref{eq:metriquebrane}--\ref{eq:tenseurelectrique}) take
the same form, but the powers are now given by:
\begin{equation}
n_x = -\frac{2 (d_\perp-2)\kappa^2}{(p+d_\perp-1)\, {\cal A}_{WZ}^2}
\ \ \ \
n_y =   \frac{2 (p+1) \kappa^2}{(p+d_\perp-1)\, {\cal A}_{WZ}^2}
\ \ \ \
n_\Phi = \frac{\alpha_p}{{\cal A}_{WZ}^2} \ ,
\end{equation}
and the relation between the Wess--Zumino coupling and the dilaton
coupling becomes:
\begin{equation}
{\cal A}_{WZ}^2 = 2 \kappa^2 \frac{(p+1)(d_\perp-2)}{(p+d_\perp-1)}
+ \frac{\alpha_p^2}{2} \ .
\end{equation}
The function $H$ is harmonic in the transverse space:
\begin{equation}
\Delta_\perp H \equiv
\delta^{IJ}\frac{\partial^2 H}{\partial y^I \partial y^J}
= -\sfrac12 {\cal A}_{WZ}^2 M^{p+1}
e^{-\alpha_p \phi_\infty /2}\ \delta^{d_\perp} (y) \ ,
\end{equation}
The rotational invariant solution is
\begin{equation}
	\label{eq:H}
H =  l + \frac{{\cal A}_{WZ}^2 M^{p+1}}{2(d_\perp-2)\Omega_{d_\perp-1}} \
e^{-\alpha_p \phi_\infty /2}\ \frac{1}{r^{d_\perp-2}}
\end{equation}
where $l$ is an arbitrary constant and
$\Omega_{d_\perp-1}$ is the volume of $S^{d_\perp-1}$.
(When $d_\perp=1$ the sphere degenerates into two points, giving
$\Omega_{0}=2$.)
The case of a brane of codimension two involves
logarithmic behavior, and we will not specify it
in the following.
Whatever the value of $l$ is, (\ref{eq:H}) gives a solution
to the equations of motion, however the solution associated to
$l=0$ has enhanced symmetry properties.
Moreover, as we will now demonstrate, when the dilaton decouples
from  the brane, the geometry of this $l=0$ solution
can be derived from effective cosmological constants.
Indeed the components of the Einstein tensor
associated with the solution (\ref{eq:H}) are
\begin{eqnarray}
	\nonumber
G^{\mu\nu} & = &
-\frac{\kappa^2}{{\cal A}_{WZ}^2}\,
\left( 1 - \frac{{\alpha_p^2}}{2\,{\cal A}_{WZ}^2 } \right)
H^{-2(n_y+1)} \,  (H^\prime)^2 g^{\mu\nu}
\\
&&  \rule{1cm}{0pt}
-\kappa^2 M^{p+1} \, H^{-(1+n_x(p+1))} e^{-\alpha_p \phi_\infty/2} \,
\frac{\delta^{d_\perp} (y)}{\sqrt{g_\perp}} \, g^{\mu\nu}
\ ;
\\
G^{IJ} & = &
-\frac{\kappa^2}{{\cal A}_{WZ}^2}\,
\left( 1 + \frac{{\alpha_p^2}}{2\,{\cal A}_{WZ}^2 } \right)
\left(2 \frac{y^I y^J}{r^2}\, e^{-2B} - g^{IJ} \right)
H^{-2(n_y+1)} \,  (H^\prime)^2
\ .
\end{eqnarray}
When the dilaton decouples, $\alpha_p=0$, implying $n_x=-1/(p+1)$ and
$n_y=1/(d_\perp-2)$, the metric can then be written as:
\begin{equation}
	\label{eq:metric}
ds^2 =
\left( \frac{r}{R_0} \right)^{2(d_\perp-2)/(p+1)}
dx^\mu \otimes dx^\nu\, \eta_{\mu\nu}
+
\left( \frac{R_0}{r} \right)^{2}
dy^I \otimes dy^J\,  \delta_{IJ}  \ .
\end{equation}
with
\begin{equation}
R_0 M = \left( \frac{p+1}{(p+d_\perp-1) \Omega_{d_\perp-1}} \right)^{1/(d_\perp-2)}
e^{-\alpha_p \phi_\infty/(2d_\perp-4)}  \ .
\end{equation}
This is the geometry of $AdS_{p+2} \times S^{d_\perp-1}$;
$R_0$ is the radius of the sphere and it is related to the radius
of the $AdS$ space by $R_0 = R_{AdS} (d_\perp-2)/(p+1)$.
The expression of the Einstein tensor simplifies to:
\begin{eqnarray}
\label{eq:Gmn}
G^{\mu\nu} & = &
-\kappa^2 \left(
\Lambda_{bk}^\circ + \Lambda_{br}^\circ \frac{\delta^{d_\perp} (y)}{\sqrt{g_\perp}} \right)
 g^{\mu\nu}  \ ;
\\
G^{IJ} & = &
-\kappa^2 \Lambda_{bk}^\circ \left(
2 \frac{y^Iy^J}{R_0^2} -g^{IJ} \right) \ ;
\end{eqnarray}
where the constants $\Lambda_{bk}^\circ$ and $\Lambda_{br}^\circ$ are given by:
\begin{eqnarray}
\Lambda_{br}^\circ &=& M^{p+1} e^{-\alpha_p \phi_\infty/2}\, ;
\nonumber\\
	\label{eq:cstcosmoN}
\Lambda_{bk}^\circ &=&
\frac{d_\perp-2}{2}\,
\left( \frac{p+d_\perp-1}{p+1} \right)^{d_\perp/(d_\perp-2)} \,
\Omega_{d_\perp-1}^{2/(d_\perp-2)}\,
M^{p+d_\perp+1} e^{\alpha_p \phi_\infty/(d_\perp-2)}
\ .
\end{eqnarray}
What allows us to interpret them as effective cosmological constants
is the fact that the metric (\ref{eq:metric}) is actually a solution to
the Einstein equations derived from a generalized RS action:
\begin{equation}
	\label{eq:RSeffective}
\action = \int d^{p+1}x\, d^{d_\perp}y \, \sqrt{|g|}
\left( \frac{\courbure}{2\kappa^2}
- \Lambda_{bk} \left( g_\perp \left(\frac{r}{R_0} \right)^{2d_\perp} \right)^{-1}
\left( \frac{R}{R_0} \right)^2
- \Lambda_{br}\, \frac{\delta^{d_\perp} (y)}{\sqrt{g_\perp}} \right)
\ ;
\end{equation}
where $R$ is defined by $R^2 = y^I y^J g_{IJ}$.
It is noteworthy that when the metric in the transverse space is integrated out, {\it i.e.},
fixed to the solution of its equation of motion,
the action (\ref{eq:RSeffective}) reduces to the one introduced by RS.

In the expression (\ref{eq:cstcosmoN}), we notice that even if the
power $n_a=(d_\perp-2)/(p+1)$ is positive, the cosmological constant on
the brane is positive. Along the discussion of the section 2, this would
not be the case with only one extra
dimension, but when $d_\perp>1$ the extra transverse dimensions that
live on the sphere also contribute to the singularity in the Einstein
tensor and modify the singularity coming from the $AdS$ part of the
space.  Nevertheless, our discussion of the hierarchy problem is
unaffected by the spherical extra dimensions and thus a positive power
$n_a$ is undesirable as regards the  gauge hierarchy problem, since it
implies that the integral for the 4D effective Planck mass diverges.
However a positive power $n_a$ naturally generates a gauge coupling
unification along the lines of the scenario proposed in
\cite{RSunification}.


Just as in the case of codimension one, the effective cosmological
constants are related to the charge $Q_e$ associated to the electric
current (\ref{eq:electriccurrent}):
\begin{equation}
\Lambda_{bk} \sim Q_e^{2/(2-d_\perp)} \ \ \ \
\mbox{and} \ \ \ \
\lambda \sim Q_e
\ ,
\end{equation}
which leads to their quantization in the multibrane configuration:
$\Lambda_{br}$ goes to
$N \Lambda_{br}^\circ $ and $\Lambda_{bk}$ goes to
$N^{2/(2-d_\perp)}\,\Lambda_{bk}^\circ\,$.

Not only does going to higher codimension brane-universes cure the
problem of the imaginary Wess--Zumino term, but they can also be more
easily embedded in a superstring framework.  As we have seen, the
theory defined by $\action_{eff}+\action_P$ admits a $p$-brane solution
only when the two couplings $\alpha_p$ and ${\cal A}_{WZ}$ are related
by eq.\ (\ref{eq:WZ}) that defines a line in the parameter space. On
this line, one point, $\alpha_p=0$, has an effective description in
terms of cosmological constants in the bulk and on the brane. On the
other hand, there is another point where the $p$-brane is
supersymmetric, the couplings taking their stringy values.  The two
points may coincide as it is the case for the D-3 brane in type {\it
IIB} theory or for the branes of $M$-theory since they do not couple to
the dilaton.  At this stage, it would be interesting to incorporate in
the field theoretical analysis of RS some stringy corrections to the
supergravity action, like quadratic terms in curvarture, for instance,
since they can modify the spectrum of the Kaluza--Klein graviton's
excitations.

\section{Discussion}

In this work we have presented solutions to the coupled equations for
branes in $d_\perp$ extra dimensions and the low energy bosonic states 
 of supergravity or superstring theories.  The goal was to
reproduce the effective stress-energy tensor needed for the
Randall-Sundrum solution which uses gravitational trapping to solve the
weak scale hierarchy problem.  Let us summarize the results.  

\subsubsection*{Decoupled dilaton regime}

Regardless of the dimensionality of the tranvserse space, we find that
the stress-energy tensor takes a simple form only in the limit that the
dilaton field decouples from the brane.  Then there are three cases:

$d_\perp = 1$.  It is necessary to go to an unphysical value of the Wess-Zumino
coupling, ${\cal A}_{WZ}^2 < 0$, to obtain a solution, which does however then
yield exactly the bulk and brane cosmological constants needed for the RS proposal.

$d_\perp = 2$.  This appears to be an uninteresting case, because ${\cal A}_{WZ}$
is forced to vanish, leading to trivial solutions.

$d_\perp > 2$.  We now find solutions with positive $\Lambda_{bk}$ and
physically acceptable values ${\cal A}_{WZ}^2 > 0$ for the Wess-Zumino
coupling. Some of the solutions are supersymmetric and are identified
as the usual branes of string theories.
 The bulk energy term looks conventional (constant) in the
brane components of $T_{\mu\nu}$, but it has a mild dependence on the
bulk coordinates in the tranvserse components,  $T_{IJ}$.  The warp
factor $a(Y)$ goes like $\exp(+{\rm const}|Y|)$ in coordinates where
$Y$ represents the physical distance from the brane in the bulk ($ {\rm
const} >0$).  Therefore the solution cannot be advocated to explain the
hierarchy between the Planck and electroweak scales.  This is in
qualitative agreement with the $d_\perp=2$ solution recently found in
ref.\ \cite{ChPz}.  It would therefore appear that the RS solution to
the hierarchy problem works only in the case of a single extra
dimension\footnote{Numerical solutions which we have found in the case
of $d_\perp=2$ also support this conclusion.}, or in the case of
several intersecting branes of codimension one.  On the other hand, as
shown in ref. \cite{RSunification}, despite infinitely large extra
dimensions, gauge coupling unification can naturally arise as a result
of the anomaly associated with the rescaling of the wave functions on
the brane.  Moreover the presence of the spherical extra dimensions can
help to cure some phenomenological puzzles which occur when there is
only one transverse dimension, such as electroweak symmetry breaking
and obtaining small enough neutrino masses \cite{RSunification}.

\subsubsection*{Coupled dilaton regime}

It is interesting to also consider the solutions where the dilaton does
not decouple from the brane.  The bulk energy is no longer constant in
these solutions, so the resulting stress-energy tensor does not have the
simple form proposed by RS.  Nevertheless, these solutions are equally
acceptable and may have interesting physical consequences.

$d_\perp = 1$.  It is now possible to have a real-valued Wess-Zumino
coupling and in this regime the metric develops a horizon at a finite
distance from the brane.


$d_\perp = 2$.  The solutions are no longer trivial, but have a
logarithmic dependence on the bulk coordinate.  We have not studied this
special case in detail.

$d_\perp > 2$.  The term in $T_{\hmu\hnu}$ which looked like a bulk
cosmological constant when the dilaton coupling vanished now has
nontrivial spatial dependence in the bulk.  Such behavior has recently
been proposed as a condition for avoiding the generic problem of the
incorrect Friedmann equation for the expansion of the brane
\cite{KKOP}.  In the latter, complicated and {\it a priori} unmotivated
expressions for the dependence of $T_{55}$ on $y$ were derived using
the requirement of correct cosmological expansion.  Although we have
not yet found inflationary solutions in the present
context, it would be interesting to do so in order to check whether the
$y$ dependence of $T_{55}$ advocated in ref. \cite{KKOP} can be
justified by the presence of nontrivial dilaton fields.

\section*{Appendix: the boundary of an anti-de Sitter space}

An anti-de Sitter space of dimension  $p+2$ can be seen as a hypersurface
embedded in a flat space of signature (2,$p+1$).
Let $x^\hmu$, $\hmu=0\ldots p+2$, be some coordinate system in this embedding space.
The anti-de Sitter space of radius $R$ is defined by the equation:
\beq
	\label{AdS}
x^\hmu x_\hmu \equiv -x^0x^0 + x^1x^1 + \ldots x^{p+1} x^{p+1} - x^{p+2} x^{p+2} = - R^2
\eeq
and the metric on $AdS$ is the embedding metric.
In a convenient system of coordinates defined by
\beq
	\label{AdScoord}
X^\mu = \frac{R}{x^{p+1}+x^{p+2}}\, x^\mu \ , \mu=0 \dots p, \ \ \
\mbox{and } \ \ \ r= x^{p+1}+x^{p+2}\ ,
\eeq
the embedding metric factorizes:
\beq
ds^2 = \left( \frac{r}{R} \right)^2 \eta_{\mu\nu} \, dX^\mu \otimes dX^\nu
+ \left( \frac{R}{r} \right)^2 dr \otimes dr \ .
\eeq
The boundary of $AdS$ is  the set of points that satisfies equation
(\ref{AdS}) at the infinity of the flat space.
More precisely, we can rescale the coordinates
$x^\hmu \to x^{\prime\, h\mu}=\lambda x^\hmu$ and consider the limit $\lambda \to \infty$.
The boundary is thus defined by the projective equations
\begin{eqnarray}
&&
-x^{\prime\, 0}x^{\prime\, 0} + x^{\prime\, 1}x^{\prime\, 1}
+ \ldots + x^{\prime\, p+1} x^{\prime\, p+1} - x^{\prime\, p+2} x^{\prime\, p+2} = 0
\\
&&
x^{\prime\, \hmu} \sim \rho x^{\prime\, \hmu} \ \ \ \mbox{with }
\rho \in \Reel \setminus \{ 0 \}
\ ,
\end{eqnarray}
which clearly describe $S^1 \times S^{p}$.  In the system of coordinates
(\ref{AdScoord}), the set of solutions to the boundary equations has two
disconnected pieces: the first one is associated with $r^\prime \not =0$,
which is sent to $r = \infty$ by the rescaling,
 and it corresponds to a Minkowski space of dimension $p+1$ spanned by
$x^0\ldots x^{p}$; the second piece is associated with $r^\prime=0$,
{\it i.e.} $r=0$, and corresponds to the union of a point and $\Reel
\times S^{p-1}$.

\section*{Acknowledgements}

C.G. would like to thank E. Dudas and J. Mourad for stimulating discussions
and is grateful to the Service de Physique Th\'eorique, CEA Saclay where
this work was initiated.
This work was supported in part by the Director, Office
of Energy Research, Office of High Energy and Nuclear Physics, Division of
High Energy Physics of the U.S. Department of Energy under Contract
DE-AC03-76SF00098 and in part by the National Science Foundation under
grant PHY-95-14797.


\end{document}